\documentclass{ws-ijmpa}

\usepackage[super]{cite}
\usepackage{xcolor}
\usepackage[verbose,hypertexnames=false]{hyperref}
\usepackage{tikz}
\hypersetup{colorlinks=false,allbordercolors=blue,pdfborderstyle={/S/U/W 1}}

\usetikzlibrary{external}
\tikzset{external/system call={pdflatex \tikzexternalcheckshellescape 
		-halt-on-error
		-interaction=batchmode 
		-jobname "\image" "\texsource"
		&& pdftops -eps "\image.pdf"}}
\begin{document}

\markboth{Jiakang Bao}{Crystals and Double Quiver Yangians}

%
\catchline{}{}{}{}{}
%

\title{An Overview of Crystals and Double Quiver Yangians}

\author{Jiakang Bao}

\address{Graduate School of Physics, University of Tokyo, Tokyo 113-0033, Japan\\
jiakang.bao@phys.s.u-tokyo.ac.jp}

\maketitle

\begin{abstract}
In this review, we summarize the recent progress on the crystal melting models and the quiver algebras regarding the BPS counting. We shall consider the constructions of crystals for generic quivers and discuss the so-called double quiver Yangians/algebras. This is an invited review for International Journal of Modern Physics A.
\end{abstract}


\ccode{PACS numbers: 11.15.$-$q, 11.30.Pb}

\section{Introduction}\label{intro}
Counting the Bogomol'nyi-Parasad-Sommerfield (BPS) states has been a valuable benchmark for understanding quantum field theories (QFTs) and string theory. In the setting of Type IIA string theory on toric Calabi-Yau (CY) threefolds, the algberas of the BPS spectra \cite{Harvey:1995fq,Harvey:1996gc} were constructed in Ref.~\refcite{Li:2020rij} known as the quiver Yangians.

The story was then extended to the shifted quiver Yangians \cite{Galakhov:2021xum} to incorporate more general framings and open BPS states. The trigonometric and elliptic counterparts of the quiver Yangians can also be constructed in a similar manner, indicating the elliptic-trigonometric-rational hierarchy realized by 3d/2d/1d QFTs with 4 supercharges. Together, they are called the quiver BPS algebras \cite{Galakhov:2021vbo}. More generally, it could be possible to consider the quiver BPS algebras associated to Riemann surfaces with higher genera and generalized cohomology theories \cite{Galakhov:2021vbo,Galakhov:2023aev,Li:2023zub}.

The basic idea of quiver Yangians/BPS algebras is that they admit the crystals as their representations. The crystal melting model provides a nice combinatorial way to encode the BPS spectra, where the BPS states formed by the D-brane bound states are in one-to-one correspondence with the configurations obtained from the quivers \cite{Okounkov:2003sp,Szendroi:2007nu,Chuang:2009crq,Ooguri:2009ijd,Ooguri:2009ri}.

The quiver Yangians/BPS algebras have been extensively studied in the literature. They have well-defined coproducts \cite{guay2018coproduct,ueda2023construction,kodera2022coproduct,Noshita:2021ldl,Noshita:2021dgj,Bao:2022jhy}. This can be further applied to the Bethe/gauge correspondence \cite{Nekrasov:2009uh,Nekrasov:2009ui} with the quiver BPS algebras on the gauge side \cite{feigin2015quantum,feigin2017finite,feigin2022quantum,Litvinov:2020zeq,Litvinov:2021phc,Chistyakova:2021yyd,Kolyaskin:2022tqi,Bao:2022fpk,Galakhov:2022uyu}. Moreover, they enjoy many other properties, such as isomorphisms under Seiberg duality (at least for a special family of the algebras) \cite{Bao:2022jhy,bezerra2021braid,Bao:2023kkh,Bao:2023ece} and connections to $\mathcal{W}$-algebras \cite{feigin2021deformations,Harada:2021xnm,ueda2022affine,kodera2022coproduct,Bao:2022jhy,ueda2023example,ueda2024guay,ueda2024coproduct,ueda2024affine1,ueda2024affine2,ueda2024two} (either rectangular or non-rectangular)\footnote{In parallel, the BPS counting problem should also be related to the gauge origami system and hence the quiver W-algebras \cite{Kimura:2015rgi,Kimura:2016dys,Koroteev:2019byp,Kimura:2023bxy,Kimura:2024xpr,Kimura:2024osv,Noshita:2025bzg,Kimura:2025lfo} although to the best of the author's knowledge, whether the quiver W-algebras are the $\mathcal{W}$-algebras in the mathematical sense and how the quiver BPS algebras and the quiver W-algebras are related precisely are still not clear.}. Readers are also referred to Ref.~\refcite{feigin2011quantum,Feigin:2010qea,feigin2012quantum,schiffmann2013cherednik,feigin2013representations,feigin2016branching,tsymbaliuk2017affine,bezerra2021quantum,Noshita:2022dxv,feigin2024commutative,Matsuo:2023lky,jimbo2024combinatorial,feigin2024remarks,Konno:2021zvl,Konno:2023znm,Konno:2024olw,Galakhov:2023mak,Galakhov:2023mpc,Galakhov:2024mbz,Galakhov:2024bzs,laurie2025tensor} for more discussions on this topic.

It is then natural to ask if the quiver BPS algebras can be extended to the non-toric cases and if there is a similar construction for theories with 2 supercharges. In this short review, we shall summarize the recent progress regarding these questions as studied in Ref.~\refcite{Bao:2024ygr,Bao:2025hfu}. The review would be self-contained, but interested readers are also referred to Ref.~\refcite{Yamazaki:2008bt} and to Ref.~\refcite{Yamazaki:2010fz} for comprehensive notes on brane tilings and on crystal melting models respectively. For the 2d $\mathcal{N}=(0,2)$ quiver gauge theories, they are concisely reviewed in Ref.~\refcite{Franco:2022iap,Franco:2024lxs}. For the quiver BPS algebras for toric CY threefolds, a succinct summary can be found in Ref.~\refcite{Yamazaki:2022cdg}.

\section{Crystal Melting and Jeffrey-Kirwan Residues}\label{crystal}
It is well-known that for toric Calabi-Yau threefolds, the correponding quiver theories \cite{Feng:2000mi,Feng:2002fv,kenyon2003introduction,Feng:2004uq,Hanany:2005ve,Franco:2005rj,Hanany:2005hq,Feng:2005gw,Yamazaki:2008bt,Gulotta:2008ef} have their BPS states encoded by the combinatorial structure known as the crystal melting models. This was first discovered for the $\mathbb{C}^3$ case in Ref.~\refcite{Okounkov:2003sp}, later for the conifold case in Ref.~\refcite{Szendroi:2007nu,Chuang:2009crq}, and was generalized to any toric quivers in Ref.~\refcite{Ooguri:2009ijd}. Recently, the crystal melting models were generalized to toric Calabi-Yau fourfolds in Ref.~\refcite{Galakhov:2023vic,Franco:2023tly,Bao:2024ygr}. Here, we shall construct the crystals using the Jeffrey-Kirwan residue formula \cite{Bao:2024ygr,Bao:2025hfu} (see also Ref.~\refcite{Nekrasov:2017cih,Nekrasov:2018xsb,Nekrasov:2023nai}).

\subsection{The Jeffrey-Kirwan residue formula}\label{JKres}
Let us first recall the Jeffrey-Kirwan residue formula \cite{jeffrey1995localization} that can be used to compute the BPS partition functions following Ref.~\refcite{Benini:2013nda,Benini:2013xpa,Hwang:2014uwa,Cordova:2014oxa,Hori:2014tda}. Given a theory with gauge group $G$, denote its Weyl group as $\mathcal{W}$. The BPS index can be computed as
\begin{equation}
	\mathcal{I}=\frac{1}{|\mathcal{W}|}\sum_{\bm{u}^*\in\mathfrak{M}^*_\text{sing}}\text{JK-Res}_{\bm{u}=\bm{u}^*}(\bm{Q}(\bm{u}^*),\eta)Z_\text{1-loop}(\bm{\epsilon},\bm{u})\;,\label{JKformula}
\end{equation}
where the Cartan subalgebra of the gauge (resp.~flavour) symmetry is parametrized by the chemical potentials $\bm{u}=\{u_i\}_{i=1}^N$ (resp.~$\bm{\epsilon}=\{\epsilon_i\}_{i=1}^F$) with $N$ (resp.~$F$) being the rank of the gauge (resp.~flavour) group. To understand this formula, we need to explain the remaining notations in \eqref{JKformula}.

\paragraph{The one-loop determinant} Let us first spell out the expression of the one-loop determinant $Z_\text{1-loop}$. It would be convenient to write
\begin{equation}
	\zeta(z)=\begin{cases}
		\frac{\text{i}\,\theta(\tau,z)}{\eta(\tau)}\;,&\text{elliptic},\\
		2\text{i}\,\sin(z)\;,&\text{trigonometric},\\
		z\;,&\text{rational}.
	\end{cases}
\end{equation}
Here, we have used the elliptic functions
\begin{equation}
	\eta(\tau)=q^{1/24}\prod_{k=1}^\infty\left(1-q^k\right)\;,\quad\theta_1(\tau,z)=-\text{i}\,q^{1/8}y^{1/2}\prod_{k=1}^\infty\left(1-q^k\right)\left(1-yq^k\right)\left(1-y^{-1}q^{k-1}\right)\;,
\end{equation}
where $q=\text{e}^{2\pi\text{i}\tau}$ and $y=\text{e}^{2\pi\text{i}z}$.

For theories with four supercharges, the one-loop determinant factorizes into the contributions from the vector multiplets $V$ and the chiral multiplets $\chi$, namely $Z_\text{1-loop}=\prod\limits_VZ_V\prod\limits_\chi Z_\chi$, where
\begin{equation}
	Z_V(\bm{u})=\xi_{\mathcal{N}=4}\prod_{\alpha\in\Phi(G)}\frac{-\zeta(\alpha(\bm{u}))}{\zeta(\alpha(\bm{u})+\epsilon)}\;,\quad Z_\chi(\bm{u})=\prod_{\rho\in\mathtt{R}}\frac{-\zeta(\rho(\bm{u})+\epsilon-\epsilon_\chi)}{\zeta(\rho(\bm{u})-\epsilon_\chi)}\;.
\end{equation}
Here, $\Phi(G)$ is the root system of $G$, and $\mathtt{R}$ is (the weight decomposition of) the representation of the chiral multiplet (excluding the zero weights). The $\text{U}(1)$ symmetry charge $\epsilon_\chi$ of the chiral multiplet is a linear combination of $\epsilon_i$, and $\epsilon=\sum\limits_i\epsilon_i$. The prefactor $\xi_{\mathcal{N}=4}$ reads
\begin{equation}
	\xi_{\mathcal{N}=4}=\begin{cases}
		\left(-\frac{\eta(\tau)^3}{\text{i}\,\theta_1(\tau,\epsilon)}\right)^N\prod\limits_{i=1}^N\text{d}(2\pi\text{i}u_i)\;,&\text{elliptic},\\
		\left(-\frac{1}{2\text{i}\,\sin(\pi\epsilon)}\right)^N\prod\limits_{i=1}^N\text{d}(2\pi\text{i}u_i)\;,&\text{trigonometric},\\
		\left(-\frac{1}{\epsilon}\right)^N\prod\limits_{i=1}^N\text{d}u_i\;,&\text{rational}.
	\end{cases}
\end{equation}

For theories with two supercharges, the one-loop determinant factorizes into the contributions from the vector multiplets $V$, the chiral multiplets $\chi$ and the Fermi multiplets $\Lambda$, namely $Z_\text{1-loop}=\prod\limits_VZ_V\prod\limits_\chi Z_\chi\prod\limits_\Lambda Z_\Lambda$, where
\begin{equation}
	Z_V(\bm{u})=\xi_{\mathcal{N}=2}\prod_{\alpha\in\Phi(G)}(-\zeta(\alpha(\bm{u})))\;,\quad Z_\chi(\bm{u})=\prod_{\rho\in\mathtt{R}}\frac{1}{\zeta(\rho(\bm{u})-\epsilon_\chi)}\;,\quad Z_\Lambda(\bm{u})=\prod_{\rho\in\mathtt{R}}(-\zeta(\rho(\bm{u})-\epsilon_\Lambda))\;.
\end{equation}
The prefactor $\xi_{\mathcal{N}=2}$ reads
\begin{equation}
	\xi_{\mathcal{N}=2}=\begin{cases}
		\eta(\tau)^{2N}\prod\limits_{i=1}^N\text{d}(2\pi\text{i}u_i)\;,&\text{elliptic},\\
		\prod\limits_{i=1}^N\text{d}(2\pi\text{i}u_i)\;,&\text{trigonometric},\\
		\prod\limits_{i=1}^N\text{d}u_i\;,&\text{rational}.
	\end{cases}
\end{equation}

We remark that the limit $\epsilon\rightarrow0$ for theories with four supercharges gives the unrefined index/partition function. However, this limit can only be taken after obtaining the refined index/partition function. Otherwise, taking $\epsilon\rightarrow0$ before the evaluation of the JK residue would make $Z_\text{1-loop}$ trivial as the numerator and the denominator would cancel each other. For theories with two supercharges, it turns out that one should already take $\epsilon=0$ in $Z_\text{1-loop}$ before evaluating the JK residue. In fact, it is still not clear whether/how one could refine the index/partition function.

\paragraph{The space \texorpdfstring{$\mathfrak{M}$}{M}} In the JK residue formula, the sum is over $\bm{u}^*\in\mathfrak{M}^*_\text{sing}$. Consider the space $\mathfrak{M}=\mathfrak{h}_\mathbb{C}/\mathtt{Q}^\vee$, where $\mathfrak{h}$ is the Cartan subalgebra of $\mathfrak{g}=\text{Lie}(G)$ and $\mathtt{Q}^\vee$ is the coroot lattice. Then the poles of $Z_\text{1-loop}$ give rise to the hyperplanes $H_i=\{Q_i(\bm{u})+\dots=0\}\subset\mathfrak{M}$. For example, we have $Q_i(\bm{u})-\epsilon_\chi=0$ with $Q_i=\rho$ from the chiral multiplet with weights $\rho$ in the representation $\mathtt{R}$.

Although we have not yet specified what the JK residue is, it is some residue as the name suggests. To compute the residue, we take the union $\mathfrak{M}_\text{sing}$ of the hyperplanes $H_i$. Then $\mathfrak{M}^*_\text{sing}$ is the set of isolated points where at least $N$ linearly independent hyperplanes meet. In this review, we shall always assume that the number of hyperplanes meeting at $\bm{u}^*$ is exactly $N$. We will briefly comment on the cases with more than $N$ hyperplanes later.

\paragraph{The Jeffrey-Kirwan residue} The JK residue is given by \cite{jeffrey1995localization}
\begin{equation}
	\text{JK-Res}\frac{\text{d}Q_{i_1}(\bm{u})}{Q_{i_1}(\bm{u})}\wedge\dots\wedge\frac{\text{d}Q_{i_N}(\bm{u})}{Q_{i_N}(\bm{u})}=\begin{cases}
		\text{sgn}(\det(Q_{i_1},\dots,Q_{i_N}))\;,&\eta\in\text{Cone}\left(Q_{i_j}\right)\;,\\
		0\;,&\text{otherwise}.
	\end{cases}
\end{equation}
Here, the covector $\eta\in\mathfrak{h}^*$ picks out the set of hyperplanes that would give non-trivial results in the JK residue. The cone is given by
\begin{equation}
	\text{Cone}\left(Q_{i_j}\right):=\left\{\sum\limits_{j=1}^Na_jQ_{i_j}\Bigg|a_j\geq0\right\}\;.
\end{equation}
In general, the covector $\eta$ should also satisfy $\eta\notin\text{Cone}_\text{sing}\left(Q_{i_1},\dots,Q_{i_{N-1}}\right)$, where $\text{Cone}_\text{sing}\left(Q_{i_1},\dots,Q_{i_{N-1}}\right)$ denotes the union of all the cones generated by $N-1$ elements in $\{Q_{i_j}\}$. The JK residue can be rewritten as
\begin{equation}
	\text{JK-Res}\frac{\text{d}u_1\wedge\dots\wedge\text{d}u_N}{Q_{i_1}(\bm{u})\dots Q_{i_N}(\bm{u})}=\begin{cases}
		\frac{1}{|\det(Q_{i_1},\dots,Q_{i_N})|}\;,&\eta\in\text{Cone}\left(Q_{i_j}\right)\;,\\
		0\;,&\text{otherwise}\;.
	\end{cases}
\end{equation}

There is an equivalent constructive definition given in Ref.~\refcite{szenes2003toric}. Instead of spelling out the detaile prescription, we shall only mention the basic idea here. In short, the JK residue can be computed as a sum of iterated residues. The poles are still specified by the covector $\eta$. When we discuss the crystals in the next subsection, this corresponds to adding atoms to the crystal in different orders. Nevertheless, a crsytal configuration would always give the same contribution regardless of the orders of adding atoms. This is consistent with the equivalence of the two definitions of JK residues. In particular, summing different orders of adding atoms is cancelled by the factor $1/|\mathcal{W}|$.

\paragraph{Example} Let us illustrate this with an example. Take the $\mathbb{C}^3$ quiver
\begin{equation}
	\includegraphics[width=3cm]{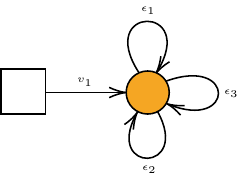}\label{C3quiver}
\end{equation}
with superpotential $W=\text{tr}(X_1X_2X_3-X_1X_3X_2)$, where $X_i$ are the three adjoint chirals transforming under the gauge group $\text{U}(N)$. The three chirals $X_\alpha$ have weights $\epsilon_\alpha$ satisfying $\epsilon_1+\epsilon_2+\epsilon_3=0$. There is also a square/framing node representing a $\text{U}(1)$ flavour symmetry. The arrow with weight $v_1$ transforms under the fundamental (resp.~anti-fundamental) of the gauge (resp.~flavour) node. For simplicity, let us take the rational case. Then the vector multiplet gives the contribution
\begin{equation}
	Z_V(\bm{u})=\left(-\frac{1}{\epsilon}\right)^N\left(\prod_{i\neq j}^N\frac{-(u_i-u_j)}{u_i-u_j+\epsilon}\right)\left(\prod_{i=1}^N\text{d}u_i\right)\;,
\end{equation}
where $\epsilon=\epsilon_1+\epsilon_2+\epsilon_3$. The adjoint $X_\alpha$ has the factor
\begin{equation}
	Z_{X_\alpha}(\bm{u})=\left(\frac{-(\epsilon-\epsilon_\alpha)}{-\epsilon_\alpha}\right)^N\prod_{i\neq j}^N\frac{-(u_j-u_i+\epsilon-\epsilon_\alpha)}{u_j-u_i-\epsilon_\alpha}\;.
\end{equation}
For the chiral connected to the framing node, we have
\begin{equation}
	Z_\mathfrak{f}=\prod_{i=1}^N\frac{-(u_i+\epsilon-v_1)}{u_i-v_1}\;.
\end{equation}
The one-loop determinant is then the product of these contributions. From these expressions, it is clear that we should not only take $\epsilon\rightarrow0$ before evaluating the residues.

The hyperplanes have covectors $Q$ given by $e_i-e_j$ and $e_i$, where $e_i$ denotes the vector with 1 at the $i^\text{th}$ entry (and 0 at other entries). One may check that all the poles from the vector multiplet get cancelled by some factors in the numerator of $Z_\text{1-loop}$. Therefore, only the matter multiplets would contribute poles to the residues. The JK residue is then the sum over all possible combinations of $N$ such covectors with $\eta$ inside the cone these $N$ covectors span.

For concreteness, let us take $\eta=(1,1,\dots,1)$. It turns out that the index computed from the JK residue for each $N$ can be encapsulated in the following generating function \cite{maulik2006gromovI,maulik2006gromovII,zhou2018donaldson}:
\begin{equation}
	\mathcal{Z}=M(-t)^{-\frac{(\epsilon_1+\epsilon_2)(\epsilon_1+\epsilon_3)(\epsilon_2+\epsilon_3)}{\epsilon_1\epsilon_2\epsilon_3}}\;,
\end{equation}
where
\begin{equation}
	M(x)=\prod\limits_{k=1}^\infty\frac{1}{(1-x^k)^k}
\end{equation}
is the MacMahon function. At order $t^N$, the coefficient gives the refined index for the $\text{U}(N)$ gauge group. In the unrefined limit $\epsilon_1+\epsilon_2+\epsilon_3=0$, this reduces to $\mathcal{Z}=M(-t)$.

\paragraph{Weights of the arrows} Given a quiver, the weights of its arrows are not always independent. Each monomial term $P$ in the superpotential gives the loop constraint:
\begin{equation}
	\sum_{I\in P}\epsilon_I=0\;.
\end{equation}
There could still be some redundancies in the parametrization since we can still change the values of $\epsilon_I$ by gauge transformations:
\begin{equation}
	\epsilon_I\rightarrow\epsilon_I+\epsilon_a\text{sgn}_a(I)
\end{equation}
for some parameters $\epsilon_a$, where $I\in a$ stands for the edges that are connected to the node $(a)$ and $\text{sgn}_a(I)$ is 1 (resp.~$-1$) is $I$ starts from (resp.~ends at) the node $(a)$. One way to eliminate such redundancies is to consider the vertex constraints:
\begin{equation}
	\sum_{I\in a}\text{sgn}_a(I)\epsilon_I=0\;.
\end{equation}
Nevertheless, it would sometimes be more convenient to keep the gauge redundancies in the calculations.

\paragraph{Fermi multiplets and brane brick models} In a 2d $\mathcal{N}=(0,2)$ theory\cite{Witten:1993yc,Garcia-Compean:1998sla,Gadde:2013lxa,Kutasov:2013ffl,Franco:2015tna,Franco:2015tya,Franco:2016nwv,Franco:2016qxh,Franco:2016fxm,Franco:2017cjj,Franco:2018qsc,Franco:2022iap,Franco:2022gvl,Franco:2022isw,Kho:2023dcm,Franco:2023tyf,Franco:2024lxs,Carcamo:2025het}, the superpotential contains term of the form $\Lambda J(X)$ for a Fermi multiplet $\Lambda$. The $J$-term $J(X)$ is a holomorphic of the chiral fields collectively denoted as $X$ here. Equivalently, one may replace $\Lambda$ with its conjugate $\overline{\Lambda}$, and the $J$-term is replaced by the corresponding $E$-term $E(X)$. If the theory can be represented by a quiver, this means that the Fermi multiplets are unoriented edges. The freedom of choosing $\Lambda$ or $\overline{\Lambda}$ causes ambiguity when writing down the 1-loop determinant. Readers are referred to the references listed above for more details on 2d $\mathcal{N}=(0,2)$ theories, their dynamics and constructions in terms of quivers. Here, we are interested in fixing the ``orientations'' for the Fermi multiplets.

Mathematically, the contributions from the chiral and Fermi multiplets correspond to the deformation and obstruction spaces when defining the DT invariants for (toric) Calabi-Yau fourfolds\cite{cao2013donaldson,cao2014donaldson,cao2017relative,Cao:2017swr,Cao:2020vce,Cao:2019fqq,Cao:2019tvv,Oh:2020rnj,Cao:2020huo,Monavari:2022rtf,Fasola:2023ypx,Kool:2025qou,Kimura:2025lig}. In particular, the definition of the DT4 invariants depends on the choice of the orientation of a certain real line bundle on the Hilbert scheme. Such an orientation can be chosen following the method in Ref.~\refcite{Monavari:2022rtf,Fasola:2023ypx,Kool:2025qou}. Physically, this corresponds to fixing the ``orientation'' of the Fermi multiplets, and we are now going to see how this can be done using the brick matchings at least when the brane brick model of the theory is known\footnote{A priori, there is nothing that prevents us from choosing a different orientation/sign as different choices are equivalent in the Lagrangian. However, they do not seem to be created equal as there exists some canonical one in the sense of Ref.~\refcite{Monavari:2022rtf}, and somehow it could be detected from the brick matchings. Such a choice would give the closed expressions of the partition functions of all the examples we know.}.

The brane brick model is a very useful tool to connect the 2d $(0,2)$ quiver gauge theories and Calabi-Yau fourfolds, generalizing the brane tiling/dimer model for the 4d $\mathcal{N}=1$ gauge theories and Calabi-Yau threefolds. For brevity, let us start from the brick matching matrix encoding the brick matchings. The definitions and the details of the brane brick models and the brick matchings can be found in Ref.~\refcite{Franco:2015tya}. It would be most straightforward to illustrate this process with an example. Consider the conifold$\times\mathbb{C}$ theory with two gauge nodes labelled by 1 and 2:
\begin{equation}
	\includegraphics[width=4cm]{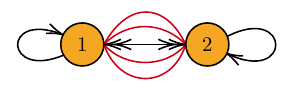}\;.
\end{equation}
The $J$- and $E$-terms are given by
\begin{equation}
	\begin{tabular}{ccc}
		& $J$ & $E$ \\
		$\Lambda^1$ & $X_{21}X_{12}Y_{21}-Y_{21}X_{12}X_{21}$ & $\Phi_{11}Y_{12}-Y_{12}\Phi_{22}$ \\
		$\Lambda^2$ & $X_{12}Y_{21}Y_{12}-Y_{12}Y_{21}X_{12}$ & $\Phi_{22}X_{21}-X_{21}\Phi_{11}$ \\
		$\Lambda^3$ & $Y_{21}Y_{12}X_{21}-X_{21}Y_{12}Y_{21}$ & $\Phi_{11}X_{12}-X_{12}\Phi_{22}$ \\
		$\Lambda^4$ & $Y_{12}X_{21}X_{12}-X_{12}X_{21}Y_{12}$ & $\Phi_{22}Y_{21}-Y_{21}\Phi_{11}$
	\end{tabular}\;.
\end{equation}
Here, $X_{ab}$, $Y_{ab}$ and $\Phi_{ab}$ are the chirals pointing from node $(a)$ to node $(b)$. For the Fermis $\Lambda$, they all connect both of the nodes and are distinguished by the superscripts. The brick matching matrix is
\begin{equation}
	P=\left( 
	\begin{array}{c|ccccc}
		 & p_1 & p_2 & p_3 & p_4 & s \\ \hline
		X_{21} & 1 & 0 & 0 & 0 & 0 \\
		X_{12} & 0 & 1 & 0 & 0 & 0 \\
		Y_{21} & 0 & 0 & 1 & 0 & 0 \\
		Y_{12} & 0 & 0 & 0 & 1 & 0 \\
		\Phi_{11} & 0 & 0 & 0 & 0 & 1 \\
		\Phi_{22} & 1 & 0 & 0 & 0 & 1 \\
	\end{array}
	\right)\;,
\end{equation}
where there are five brick matchings in this case. Let us first choose the brick matching $p_1$. The only non-zero entry in the $p_1$ column corresponds to $X_{21}$. Then the rule is to choose the $J$- and $E$-terms that contain $X_{21}$. For the $\Lambda^1$ row, the $J$ column is chosen. Therefore, we choose the orientation of $\Lambda^1$ as from node (1) to node (2) so that $\Lambda^1$ together with each monomial in the $J$-term would form a closed oriented loop\footnote{This can be seen as follows. Write the ``oriented'' $\Lambda^1$ as $\Lambda^1_{12}$. Then the subscripts of the fields should be connected end-to-end. For example, we have $\Lambda^1_{12}X_{21}X_{12}Y_{21}$ and $\Lambda^1_{12}Y_{21}X_{12}X_{21}$ here. In the quiver, this forms a closed oriented loop.}. Likewise, for $\Lambda^2$ ($\Lambda^3$, resp.~$\Lambda^4$), the $J$ ($E$, resp.~$J$) column is chosen. As a result, they are assigned orientations $2\rightarrow1$, $2\rightarrow1$ and $2\rightarrow1$ respectively.

Then in the 1-loop determinant, since the Fermis transform as the bifundamentals, they give the factors $\zeta\left(u^{(b)}_j-u^{(a)}_i-\epsilon_\Lambda\right)$. Here, the arrow is from node $(a)$ to node $(b)$, and $i$, $j$ run from 1 to their corresponding gauge node ranks.

We may also pick other brick matchings, say $s$. Then all the four Fermis have the $E$ column chosen so that the orientations are $2\rightarrow1$, $1\rightarrow2$, $2\rightarrow1$, $1\rightarrow2$ respectively. Although each individual $Z_\Lambda$ would be different for different choices of the brick matchings, the total 1-loop determinant, and hence the index, would be the same regardless of this choice. The closed expression of the partition function in the rational case was conjectured in Ref.~\refcite{Bao:2024ygr}, including the wall crossing of the second kind\cite{Aganagic:2010qr}.

\subsection{Crystals}\label{crystals}
With the JK residue formula introduced, we are ready to define the crystals that encode the information of BPS counting combinatorially. For toric Calabi-Yau threefolds and fourfolds, the crystals can be constructed from the periodic quivers, where one uplifts the quiver diagrams by one dimension, and the crystal melting picture follows the paths of the arrows in the periodic quivers \cite{Ooguri:2009ijd,Franco:2023tly}. In fact, it can be shown that the crystal structure can be determined by the pole structure in the JK residue formula \cite{Bao:2024ygr}. Now, with the JK residue formula, many aspects may be carried over to more general quivers satisfying certain constraints \cite{Bao:2025hfu}.

Here, we shall consider the quiver gauge theories with unitary gauge groups $\text{U}(N_a)$ (and also unitary flavour nodes). The vector of the gauge group ranks $\bm{N}=(N_a)$ is called the dimension vector, and the sum of the ranks will be denoted as $N$. The quiver arrows are always bifundamental or adjoint matters (or (anti-)fundamental if connected to a framing node). Moreover, the superpotential should be polynomials of these fields.

To construct the crystal, we need to vary the dimension vector $\bm{N}$, and hence the covector $\eta_{\bm{N}}$. Correspondingly, we shall fix a (possibly infinite-size) covector $\eta$ which can be truncated to $\eta_{\bm{N}}$ at a finite size.

Let us denote by $\mathfrak{U}(\bm{N},\eta)$ the set of $\bm{u}^*=\left(u^{(a)*}_i\right)\in\mathfrak{M}_\text{sing}$ such that the JK residue is non-zero at $\bm{u}^*$ (i.e., it is admissible by $\eta$ and gives non-zero residues). Each $\bm{u}^*\in\mathfrak{U}(\bm{N},\eta)$ is an isolated point where at least $N$ hyperplanes intersect. Denote the set of these hyperplanes as $\mathfrak{H}(\bm{u}^*)$. In particular, each $u^{(a)*}_i$ is a linear combination of the equivariant parameters $\epsilon_k$:
\begin{equation}
	u^{(a)*}_i\in\bigoplus_{k=1}^F\mathbb{Z}\epsilon_k
\end{equation}
for each $a\in Q_0$ and $i\in\{1,\dots,N_a\}$. Here, $Q_0$ is the set of all gauge nodes in the quiver and should not be confused with the covectors $Q_i$ that appear in the definition of the JK residue.

Define
\begin{equation}
	\mathcal{A}(\bm{u}^*):=\left\{u^{(a)*}_i\;\bigg|\;a\in Q_0,\;i=1,\dots,N_a\right\}\subset\bigoplus_{k=1}^F\mathbb{Z}\epsilon_k\;.
\end{equation}
We can collect these sets and write
\begin{equation}
	\mathcal{A}(\bm{N},\eta):=\bigcup_{\bm{u}^*\in\mathfrak{U}(\bm{N},\eta)}\mathcal{A}(\bm{u}^*)\;.
\end{equation}
We shall refer to this as the set of atoms at level $\bm{N}$. We may then write the set of atoms at level $N$:
\begin{equation}
	\mathcal{A}(N,\eta):=\bigcup_{\bm{N}:\sum\limits_aN_a=N}\mathcal{A}(\bm{N},\eta)\;.
\end{equation}
The full set of atoms is then defined as
\begin{equation}
	\mathcal{A}(\eta)=\bigcup_{\bm{N}\in\mathbb{Z}_{\geq0}^{|Q_0|}}\mathcal{A}(\bm{N},\eta)=\bigcup_{N\in\mathbb{Z}_{\geq0}}\mathcal{A}(N,\eta)\;.
\end{equation}

Consider the partial ordering $\bm{M}\leq\bm{N}$ given by $M_a\leq N_a$ for all $a\in Q_0$. Then
\begin{equation}
	\mathcal{A}(\bm{M},\eta)\subseteq\mathcal{A}(\bm{N},\eta)\;.
\end{equation}
This means that we can write the full set of atoms as a direct limit with respect to this partial ordering:
\begin{equation}
	\mathcal{A}=\lim_{\longrightarrow}\mathcal{A}(\bm{N},\eta)\;.
\end{equation}

The crystal $\mathcal{C}(\eta)=(\mathcal{A}(\eta),\mathcal{I}(\eta))$ is defined as an oriented weighted graph with the set of vertices $\mathcal{A}(\eta)$ and the set of arrows $\mathcal{I}(\eta)$. Each vertex will be called an atom denoted as $\mathfrak{a}$ corresponding to $a$. The arrows are given as follows. Suppose that there are two atoms $\mathfrak{a}$ and $\mathfrak{b}$. We draw an arrow $I$ from $\mathfrak{b}$ to $\mathfrak{a}$ if
\begin{itemize}
	\item there exists $\bm{N}\in\mathbb{Z}_{\geq0}^{|Q_0|}$ and $\bm{u}^*\in\mathfrak{U}(\bm{N},\eta)$ such that $\mathfrak{a},\mathfrak{b}\in\mathcal{A}(\bm{u}^*)$,
	\item and there exists a hyperplane of the form
	\begin{equation}
		\left\{u^{(a)}_j-u^{(b)}_i-\dots=0\right\}
	\end{equation}
	inside $\mathfrak{H}(\bm{u}^*)$.
\end{itemize}
Such an arrow will be called the chemical bond, and it has weight $\epsilon_I=\epsilon_{\mathfrak{a}}-\epsilon_{\mathfrak{b}}$.

\paragraph{No-overlap condition} Here, we need an extra condition on the crystal, which is a sufficient (but not necessary) condition to ensure that the JK residue formula gives the right counting. This is called the no-overlap condition as follows.

Suppose that we are given an atom $\mathfrak{a}$ in $\mathcal{A}(\eta)$ such that $\mathfrak{a}\in\mathcal{A}(\bm{N},\eta)$ for some $\bm{N}$. In general, this can have multiple realizations inside $\mathcal{A}(\bm{N},\eta)$ so that
\begin{equation}
	\mathfrak{a}=u^{(a)*}_i=u^{(b)*}_j
\end{equation}
for different pairs $(i,a)$ and $(j,b)$ (in other words, either $i\neq j$, or $a\neq b$ if $i=j$), but with the same $\bm{u}^*\in\mathcal{A}(\bm{N},\eta)$. If this ever happens for some $\bm{N}$, we say that the no-overlap condition is violated. Otherwise, we say that no-overlap condition is satisfied.

Intuitively, this condition states that the atoms are not allowed to overlap in the crystal, and hence the name. We require this condition so that the poles in the 1-loop determinants are all simple poles. When there are higher order poles in the one-loop determinant, this does not imply that the JK residue formula would lose its effecacy. One can still evaluate the residues for the higher order poles, or equivalently,  shift the coincident hyperplanes apart to resolve the higher order poles into multiplet simple poles. Some examples can be found in Ref.~\refcite{Benini:2013xpa}. However, for some cases, resolving the coincident hyperplanes would still not give the full correct partition function. There could be contributions from non-isolated fixed loci in addition to the fixed points\footnote{We would like to thank Taro Kimura and Henry Liu for pointing this out.}. An example was discussed in Ref.~\refcite{Nekrasov:2015wsu}. Therefore, for simplicity, we shall always assume the no-overlap condition in this review.

\paragraph{Molecules and cyclic chambers} Consider the crystal $\mathcal{C}=(\mathcal{A},\mathcal{I})$. We shall define the molecule as a finite subset $\mathcal{M}\subseteq\mathcal{A}$ such that the following condition (melting rule) is satisfied.

Suppose that $\mathfrak{a},\mathfrak{b}\in\mathcal{M}$ with an arrow $I\in\mathcal{I}$ connecting from $\mathfrak{a}$ to $\mathfrak{b}$. If $\mathfrak{b}\in\mathcal{M}$, then $\mathfrak{a}\in\mathcal{M}$.

Following the melting rule, one can add or remove atoms to get a larger or smaller molecule. This is in line with the constructive definition of the JK residue, which indicates that the partition function at level $N+1$ can be computed from the one at level $N$:
\begin{equation}
	Z_\text{1-loop}(u_1,\dots,u_{N+1})=Z_\text{1-loop}(u_1,\dots,u_N)\Delta Z(u_1,\dots,u_{N+1})\;.\label{ZZDeltaZ}
\end{equation}
However, in the JK residue formula, there is an extra input, namely $\eta$. For certain $\eta$, it could be possible that $\eta_{N+1}\in\text{Cone}(H_1,\dots,N_{N+1})$ while $\eta_N\notin\text{Cone}(H_1,\dots,H_N)$ (where the subscript $N$ of $\eta$ denotes the truncation of $\eta$ onto the first $N$ elements):
\begin{equation}
	\includegraphics[width=8cm]{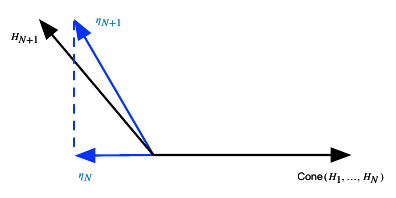}\;.
\end{equation}
Therefore, it could be possible that $\mathcal{M}$ and $\mathcal{M}+\mathfrak{a}+\mathfrak{b}$ are non-trivial contributions to the counting while $\mathcal{M}+\mathfrak{a}$ is not. In other words, a minimal step of the growth may involve multiple atoms.

We shall also refer to the molecules that correspond to the non-trivial counting as the crystal states. When a minimal step of melting is always one single atom, we say that this choice gives the cyclic chamber. In general, different $\eta$ is expected to be related by the wall crossing (of the first kind)\footnote{In 2d, as the 1-loop determinant is defined on the compact torus, the elliptic genus should be independent of $\eta$ and does not experience the wall crossing (of the first kind) \cite{Benini:2013xpa,liu2025invariance}. Therefore, we should always have the cyclic chamber in this case by choosing $\eta=(1,\dots,1)$.}. As shown in Ref.~\refcite{Bao:2025hfu}, the choice $\eta=(1,1,\dots,1)$ would always give the cyclic chamber. In the followings, we shall always take $\eta=(1,1,\dots,1)$ unless otherwise speficied\footnote{See Ref.~\refcite{Galakhov:2024foa} for some recent study on the non-cyclic chambers for the conifold case.}.

\begin{figure}[h]
	\centering
	\includegraphics[width=10cm]{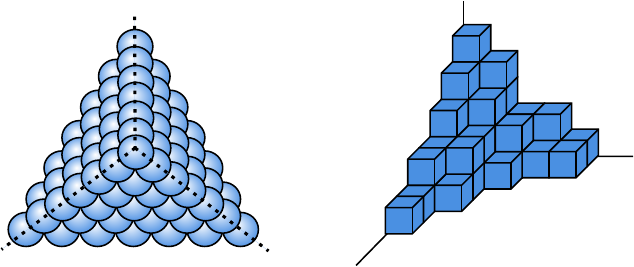}
	\caption{The crystal of the quiver \eqref{C3quiver}. The right plot is equivalent to the left one but with a view from a different angle where the plane partition nature is more clear.}\label{C3crystal}
\end{figure}

\paragraph{Example} Let us consider the example with quiver \eqref{C3quiver} discussed above. From the pole structure, the crystal states are in one-to-one correspondence with the plane partitions. This is in line with the partition function given by the MacMahon function (up to signs). An illustration can be found in Fig.~\ref{C3crystal}.
As shown in Ref.~\refcite{Bao:2024ygr}, the crystals obtained from the periodic quivers in the toric cases (either $\mathcal{N}=4$ or $\mathcal{N}=2$) can indeed be recovered from the JK residue formula\footnote{For the $\mathbb{C}^4$ case, this was first proven in Ref.~\refcite{Nekrasov:2017cih}.}.

For the toric cases and many other cases, the poles are always simple poles, and the no-overlap condition is satisfied. However, quivers such as those associated to the affine D-type and E-type Dynkin diagrams would violate the no-overlap condition. An example can be found in Ref.~\refcite{Bao:2025hfu}.

\paragraph{Modules of truncated Jacobi algebras} Let us briefly mention that the crystals can also be constructed as the modules of the truncated Jacobi algebras. In toric cases, the crystals are the modules of the Jacobi algebras\cite{Ooguri:2009ijd}, which are the path algebras quotiented by the $F$-term or $J$-/$E$-term relations. However, for non-toric cases, since the $F$-term or $J$-/$E$-term relations could be of the form
\begin{equation}
	\sum_{i=1}^nP_i=0
\end{equation}
with $n>2$ (while $n=2$ in the toric cases), where $P_i$ are monomials of the chiral fields, the relations quotiented in the path algebras should be stronger. More concretely, the crystals are the modules of the truncated Jacobi algebras
\begin{equation}
	\mathcal{J}^{\#}=\mathbb{C}Q/\langle P_1=P_2=\dots=P_n\rangle
\end{equation}
for the quivers $Q$. See Ref.~\refcite{Bao:2025hfu} for more details.

\section{Quiver Algebras}\label{algebras}
With the discussions of the JK residues and the crystals, we are ready to construct the quiver algebras that encode the information of BPS counting. For theories with 4 supercharges, the quiver BPS algebras were introduced in Ref.~\refcite{Li:2020rij,Galakhov:2021xum,Galakhov:2021vbo}. Such algebras were also derived physically from supersymmetric quantum mechanics in Ref.~\refcite{Galakhov:2020vyb}. For theories with 2 supercharges, the situations would become quite different. Nevertheless, we shall extract some algebraic structure using the JK residue method. We shall refer to the algebras constructed this way as the double quiver Yangians/algebras.

\subsection{Quiver Yangians/BPS algebras}\label{quiverbpsalgebras}
Let us first recall the definition of the quiver Yangians/BPS algebras\footnote{More precisely, the quiver Yangian is the rational quiver BPS algebra while one can also have the trigonometric and elliptic versions of the quiver BPS algebra depending on $\zeta(z)$. Nevertheless, we shall use the two names interchangeably in this review.}. There are three sets of generators $\psi^{(a)}_n$, $e^{(a)}_n$ and $f^{(a)}_n$, where $a\in Q_0$ and $n\in\mathbb{Z}$. They can be packaged into the currents $\psi^{(a)}(z)$, $e^{(a)}(z)$ and $f^{(a)}(z)$ whose expressions can be found for example in (2.17)$\sim$(2.19) in Ref.~\refcite{Galakhov:2021vbo}. The relations can be most conveniently written in terms of the currents\footnote{We shall use the capital letters to denote the exponentials of the corresponding lower case letters. For instance, $C=\text{e}^{2\pi\text{i}c}$, $Z=\text{e}^{2\pi\text{z}}$.}:
\begin{align}
	&\psi^{(a)}_{\pm}(z)\psi^{(b)}_{\pm}(w)\simeq C^{\pm\chi_{ab}}\psi^{(b)}_{\pm}(w)\psi^{(a)}_{\pm}(z)\;,\\
	&\psi^{(a)}_+(z)\psi^{(b)}_-(w)\simeq\frac{\varphi^{a\Leftarrow b}(z+c/2,w-c/2)}{\varphi^{a\Leftarrow b}(z-c/2,w+c/2)}\psi^{(b)}_-(w)\psi^{(a)}_+(z)\;,\\
	&\psi^{(a)}_{\pm}(z)e^{(b)}(w)\simeq\varphi^{a\Leftarrow b}(z\pm c/2,w)e^{(b)}(w)\psi^{(a)}_{\pm}(z)\;,\\
	&\psi^{(a)}_{\pm}(z)f^{(b)}(w)\simeq\varphi^{a\Leftarrow b}(z\mp c/2,w)^{-1}f^{(b)}(w)\psi^{(a)}_{\pm}(z)\;,\\
	&e^{(a)}(z)e^{(b)}(w)\simeq(-1)^{|a||b|}\varphi^{a\Leftarrow b}(z,w)e^{(b)}(w)e^{(a)}(z)\;,\\
	&f^{(a)}(z)f^{(b)}(w)\simeq(-1)^{|a||b|}\varphi^{a\Leftarrow b}(z,w)^{-1}f^{(b)}(w)f^{(a)}(z)\;,\\
	&\left[e^{(a)}(z),f^{(b)}(w)\right\}\simeq\delta_{ab}\left(\delta(z-w-c)\psi^{(a)}_+(z-c/2)-\delta(z-w+c)\psi^{(a)}_-(w-c/2)\right)\;.
\end{align}
Here, $\chi_{ab}$ is the chirality given by $\chi_{ab}=|a\rightarrow b|-|b\rightarrow a|$. The currents $e^{(a)}(z)$ and $f^{(a)}(z)$ are bosonic (resp.~fermionic) if there is an odd (resp.~even) number of adjoint loops on the node $(a)$ while the currents $\psi^{(a)}(z)$ are always bosonic. The bond factor is
\begin{equation}
	\phi^{a\Leftarrow b}(z)=(-1)^{|b\rightarrow a|}\frac{\prod\limits_{I\in\{a\rightarrow b\}}\zeta\left(z+\epsilon_I\right)}{\prod\limits_{I\in\{b\rightarrow a\}}\zeta\left(z-\epsilon_I\right)}\;.
\end{equation}
Moreover, we have the balanced bond factor
\begin{equation}
	\varphi^{a\Leftarrow b}(z,w)=(ZW)^{\frac{\mathfrak{t}}{2}\chi_{ab}}\phi^{a\Leftarrow b}(z-w)
\end{equation}
with
\begin{equation}
	\mathfrak{t}=\begin{cases}
		0\;, &\text{rational}\;,\\
		1\;, &\text{trigonometric/elliptic}\;,
	\end{cases}
\end{equation}
to avoid the fractional powers in the Laurent expansions for chiral quivers. The formal $\delta$-function is
\begin{equation}
	\delta(z)=\begin{cases}
		1/z\;,&\text{rational},\\
		\sum\limits_{k\in\mathbb{Z}}Z^k\;,&\text{trigonometric/elliptic}.
	\end{cases}
\end{equation}
For the rational case, $\psi^{(a)}(z):=\psi^{(a)}_+(z)=\psi^{(a)}_-(z)$, and $c=0$.

\paragraph{Crystal representations} The crystal states form a representation of the quiver BPS algebra when $c=0$. The actions of the generating currents are
\begin{align}
	&\psi^{(a)}_{\pm}(z)|\mathcal{C}\rangle=\left[\Psi^{(a)}_{\mathcal{C}}(z)\right]_{\pm}|\mathcal{C}\rangle\;,\\
	&e^{(a)}(z)|\mathcal{C}\rangle=\sum_{\mathfrak{a}\in\text{Add}(\mathcal{C})}\left(\pm\delta(z-\epsilon_{\mathfrak{a}})\left(\pm\lim\limits_{x\rightarrow\epsilon_{\mathfrak{a}}}\Psi^{(a)}_{\mathcal{C}}(x)\right)^{1/2}\right)|\mathcal{C}+\mathfrak{a}\rangle\;,\\
	&f^{(a)}(z)|\mathcal{C}\rangle=\sum_{\mathfrak{a}\in\text{Rem}(\mathcal{C})}\left(\pm\delta(z-\epsilon_{\mathfrak{a}})\left(\pm\lim\limits_{x\rightarrow\epsilon_{\mathfrak{a}}}\Psi^{(a)}_{\mathcal{C}}(x)\right)^{1/2}\right)|\mathcal{C}-\mathfrak{a}\rangle\;.
\end{align}
The charge function is
\begin{equation}
	\Psi^{(a)}_{\mathcal{C}}(z)={}^{\#}\psi^{(a)}(z)\prod_{b\in Q_0}\prod_{\mathfrak{b}\in\mathcal{C}}\varphi^{a\Leftarrow b}(z,\epsilon_{\mathfrak{b}}) \;,
\end{equation}
where ${}^\#\psi^{(a)}(z)$ is the factor coming from the framing that can be determined in the same manner as $\phi^{a\Rightarrow b}$. The notation $[\dots]_+$ (resp.~$[\dots]_-$) indicates the expansion of the function around $z=\infty$ (resp.~$z=0$). The key property of the charge function is that its poles, which are always simple, are in one-to-one correspondence with the addable and removable atoms in the crystal state. The $\pm$ signs in the actions of $e^{(a)}(z)$ and $f^{(a)}(z)$ can be determined following the prescriptions in Ref.~\refcite{Li:2020rij} and Ref.~\refcite{Galakhov:2021vbo}.

To verify that this is indeed a representation of the algebra, one needs to check that the actions are consistent with the algebra relations. For instance, consider $e^{(a)}(z)e^{(b)}(w)$ and $e^{(b)}(w)e^{(a)}(z)$ acting on some state $|\mathcal{C}\rangle$. We can compute the ratio of the coefficients for any resulting state $|\mathcal{C}+\mathfrak{a}+\mathfrak{b}\rangle$, which we shall formally write as\footnote{Notice that there could be multiple possible configurations $\mathcal{C}+\mathfrak{a}+\mathfrak{b}$.}
\begin{equation}
	``\frac{e^{(a)}(z)e^{(b)}(w)|\mathcal{C}\rangle}{e^{(b)}(w)e^{(a)}(z)|\mathcal{C}\rangle}"\;.\label{ee}
\end{equation}
It turns out that this is independent of $|\mathcal{C}\rangle$ and recovers the coefficient in the $ee$ relation.

In Ref.~\refcite{Bao:2025hfu}, it was shown that by keeping ``half" of the factors in the one-loop determinant $Z_\text{1-loop}$ in the JK residue formula in the following sense. When adding an atom to the crystal state, $\Delta Z$ in \eqref{ZZDeltaZ} contains the factors of the form
\begin{equation}
	\left(\prod_{I\in\{a\rightarrow b\}}\frac{-{\color{red} \zeta(z-\epsilon_{\mathfrak{b}}-\epsilon+\epsilon_I)}}{\zeta(z-\epsilon_{\mathfrak{b}}+\epsilon_I)}\right)\left(\prod_{I\in\{b\rightarrow a\}}\frac{-\zeta(z-\epsilon_{\mathfrak{b}}+\epsilon-\epsilon_I)}{{\color{red} \zeta(z-\epsilon_{\mathfrak{b}}-\epsilon_I)}}\right)\;.
\end{equation}
It turns out that by keeping only the red terms\footnote{We have also discarded the contributions from the vector multiplets and the constant factors in the adjoint chiral multiplets which are not shown here.}, the expression would precisely give the addable and removable atoms from the poles, and it is exactly the corresponding factor in the charge function\footnote{Notice that it is now safe to take $\epsilon\rightarrow0$.}.

\subsection{Double quiver algebras}\label{doublequiveralgebras}
For theories with 2 supercharges, one may still construct some $\Psi^{(a)}_{\mathcal{C}}(z)$ whose poles are precisely the addable and removable atoms for each crystal state. However, if we would like to have some raising and lowering operators $e^{(a)}(z)$ and $f^{(a)}(z)$ with respect to this charge function, the expressions like \eqref{ee} would depend on the states \cite{Galakhov:2023vic}. In other words, it would be difficult to extract the state-independent relations for the potential algebra generators. Moreover, the information of the BPS counting is only partly encoded in the crystals. There are non-trivial coefficients (which are functions of the fugacities, rather than just $\pm1$) in the partition functions for these theories.

Now that we have the JK residue formula to compute the partition functions, we may also try to keep the full 1-loop determinant as some charge function $\widetilde{\Psi}^{(a)}_{\mathcal{C}}(z)$. Explicitly,
\begin{equation}
	\widetilde{\Psi}^{(a)}_{\mathcal{C}}(z)={}^\#\widetilde{\psi}^{(a)}(z)\prod_{b\in Q_0}\prod_{\mathfrak{b}\in\mathfrak{\mathcal{C}}}\widetilde{\phi}^{a\Rightarrow b}(z-\epsilon_{\mathfrak{b}})\;,
\end{equation}
where
\begin{equation}
	\widetilde{\phi}^{a\Leftarrow b}(z):=\begin{cases}
		\displaystyle
		\frac{\zeta(z)\zeta(-z)}{\zeta(z+\epsilon)\zeta(-z+\epsilon)}\left(\prod\limits_{I\in\{a\rightarrow a\}}\frac{-\zeta\left(\epsilon-\epsilon_I\right)}{\zeta(-\epsilon_I)}\frac{\zeta(z+\epsilon-\epsilon_I)\zeta(z-\epsilon+\epsilon_I)}{\zeta(z-\epsilon_I)\zeta(z+\epsilon_I)}\right) \;,&b=a \;,\\
		\displaystyle \left(\prod\limits_{I\in\{a\rightarrow b\}}\frac{-\zeta(z-\epsilon+\epsilon_I)}{\zeta(z+\epsilon_I)}\right)\left(\prod\limits_{I\in\{b\rightarrow a\}}\frac{-\zeta(z+\epsilon-\epsilon_I)}{\zeta(z-\epsilon_I)}\right) \;,&b\neq a \;,
	\end{cases}
\end{equation}
for theories with 4 supercharges and
\begin{equation}
	\widetilde{\phi}^{a\Leftarrow b}(z)
	:=\begin{cases}
		\displaystyle
		\zeta(z)\zeta(-z)\frac{\prod\limits_{\Lambda\in\{aa\}}\zeta(\epsilon_{\Lambda})\zeta(z-\epsilon_{\Lambda})\zeta(z+\epsilon_{\Lambda})}{\prod\limits_{I\in\{a\rightarrow a\}}\zeta(\epsilon_I)\zeta(z-\epsilon_I)\zeta(z+\epsilon_I)}\;,&b=a\;,\\
		\displaystyle
		\frac{\prod\limits_{\Lambda\in\{ab\}}\left(-\zeta\left(\varsigma^{a\Leftarrow b}_{\Lambda}z-\epsilon_{\Lambda}\right)\right)}{\left(\prod\limits_{I\in\{a\rightarrow b\}}(-\zeta(z+\epsilon_I))\right)\left(\prod\limits_{I\in\{b\rightarrow a\}}\zeta(z-\epsilon_I)\right)}\;,&b\neq a\;,
	\end{cases}
\end{equation}
for theories with 2 supercharges. The factor ${}^\#\widetilde{\psi}^{(a)}(z)$ that comes from the framing can be determined in the same manner. Here, $\varsigma^{a\Leftarrow b}_{\Lambda}$ is the sign determined by the choice of the orientation of $\Lambda$, namely,
\begin{equation}
	\varsigma^{a\Leftarrow b}_{\Lambda}
	:=\begin{cases}
		+\;,&s(\Lambda)=b,~t(\Lambda)=a\;,\\
		-\;,&s(\Lambda)=a,~t(\Lambda)=b\;.
	\end{cases}
\end{equation}

Let us consider the following currents acting on the crystal states as
\begin{align}
	&\widetilde{\psi}^{(a)}_{\pm}(z)|\mathcal{C}\rangle=
		\left[\widetilde{\Psi}^{(a)}_{\mathcal{C}}(z)\right]_{\pm}|\mathcal{C}\rangle\;,\\
	&\widetilde{\omega}^{(a)}(z)|\mathcal{C}\rangle=\sum_{\text{Inad}\left(\Psi^{(a)}_{\mathcal{C}}(z),\eta\right)}\delta(z-\epsilon_{\mathfrak{a}})\lim_{x=\epsilon_{\mathfrak{a}}}\zeta(x-\epsilon_{\mathfrak{a}})\widetilde{\Psi}^{(a)}_{\mathcal{C}}(x)|\mathcal{C}\rangle\;,\\
	&\widetilde{e}^{(a)}(z)|\mathcal{C}\rangle=\sum_{\mathfrak{a}\in\text{Add}(\mathcal{C})}\pm\delta(z-\epsilon_{\mathfrak{a}})\left(\pm\lim_{x=\epsilon_{\mathfrak{a}}}\zeta(x-\epsilon_{\mathfrak{a}})\widetilde{\Psi}^{(a)}_{\mathcal{C}}(x)\right)^{1/2}|\mathcal{C}+\mathfrak{a}\rangle\;,\\
	&\widetilde{f}^{(a)}(z)|\mathcal{C}\rangle=\sum_{\mathfrak{a}\in\text{Rem}(\mathcal{C})}\pm\delta(z-\epsilon_{\mathfrak{a}})\left(\pm\lim_{x=\epsilon_{\mathfrak{a}}}\zeta(x-\epsilon_{\mathfrak{a}})\widetilde{\Psi}^{(a)}_{\mathcal{C}-\mathfrak{a}}(x)\right)^{1/2}|\mathcal{C}-\mathfrak{a}\rangle\;.
\end{align}
Compared to the quiver BPS algebras in the previous subsection, there are two main differences. Since the charge functions $\widetilde{\Psi}^{(a)}_{\mathcal{C}}(z)$ coming from the 1-loop determinant only contain the addable atoms but not the removable atoms, we have $\widetilde{\Psi}^{(a)}_{\mathcal{C}-\mathfrak{a}}(z)$ (rather than $\widetilde{\Psi}^{(a)}_{\mathcal{C}}(z)$) in the actions of $\widetilde{f}^{(a)}(z)$. Moreover, as $\widetilde{\Psi}^{(a)}_{\mathcal{C}}(z)$ have extra poles that are ruled out by $\eta$ in the computations of the JK residue formula, we have introduced the new currents $\widetilde{\omega}^{(a)}(z)$ to collect all these inadmissible poles.

We may then extract the relations among the currents as
\begin{align}
	&\widetilde{\omega}^{(a)}(z)\widetilde{\omega}^{(b)}(w)\simeq\widetilde{\omega}^{(b)}(w)\widetilde{\omega}^{(a)}(z) \;,\\
	&\widetilde{\psi}^{(a)}_+(z)\widetilde{\psi}^{(b)}_+(w)\simeq\widetilde{\psi}^{(b)}_+(w)\widetilde{\psi}^{(a)}_+(z) \;,\\
	&\widetilde{\psi}^{(a)}_-(z)\widetilde{\psi}^{(b)}_-(w)\simeq\widetilde{\phi}^{a\Leftarrow b}(z-w+2c)\widetilde{\phi}^{a\Leftarrow b}(z-w-2c)^{-1}\widetilde{\psi}^{(b)}_-(w)\widetilde{\psi}^{(a)}_-(z) \;,\\
	&\widetilde{\psi}^{(a)}_+(z)\widetilde{\psi}^{(b)}_-(w)\simeq\widetilde{\phi}^{a\Leftarrow b}(z-w+c)\widetilde{\phi}^{a\Leftarrow b}(z-w-c)^{-1}\widetilde{\psi}^{(b)}_-(w)\widetilde{\psi}^{(a)}_+(z) \;,\\
	&\widetilde{\psi}^{(a)}_{\pm}(z)\widetilde{\omega}^{(b)}(w)\simeq\widetilde{\phi}^{a\Leftarrow b}(z-w+c\mp c/2)^{-1}\widetilde{\phi}^{a\Leftarrow b}(z-w-c\pm c/2)\widetilde{\omega}^{(b)}(w)\widetilde{\psi}^{(a)}_{\pm}(z) \;,\\
	&\widetilde{d}^{(ba)}(z-w)\widetilde{\psi}^{(a)}_{\pm}(z)\widetilde{e}^{(b)}(w)\simeq\widetilde{\phi}^{a\Leftarrow b}(z-w+c\mp c/2)\widetilde{d}^{(ba)}\widetilde{e}^{(b)}(w)\widetilde{\psi}^{(a)}_{\pm}(z) \;,\\
	&\widetilde{d}^{(ab)}(z-w)\widetilde{\psi}^{(a)}_{\pm}(z)\widetilde{f}^{(b)}(w)\simeq\widetilde{\phi}^{a\Leftarrow b}(z-w-c\pm c/2)^{-1}\widetilde{d}^{(ab)}(z-w)\widetilde{f}^{(b)}(w)\widetilde{\psi}^{(a)}_{\pm}(z) \;,\\
	&\delta(z-w)\widetilde{\phi}^{d\Leftarrow a}(u-z-c)\widetilde{e}^{(d)}(u)\widetilde{\omega}^{(a)}(z)+\delta(u-w)\widetilde{\phi}^{a\Leftarrow d}(z-w-c)\widetilde{e}^{(a)}(z)\widetilde{\omega}^{(d)}(w)\nonumber\\
	&\simeq \delta(z-w)\widetilde{\omega}^{(a)}(z)\widetilde{e}^{(d)}(u)+\delta(u-w)\widetilde{\omega}^{(d)}(u)\widetilde{e}^{(a)}(z) \;,\\
	&\delta(z-w)\widetilde{\phi}^{d\Leftarrow a}(u-z-c)^{-1}\widetilde{f}^{(d)}(u)\widetilde{\omega}^{(a)}(z)+\delta(u-w)\widetilde{\phi}^{a\Leftarrow d}(z-w-c)^{-1}\widetilde{f}^{(a)}(z)\widetilde{\omega}^{(d)}(w)\nonumber\\
	&\simeq \delta(z-w)\widetilde{\omega}^{(a)}(z)\widetilde{f}^{(d)}(u)+\delta(u-w)\widetilde{\omega}^{(d)}(u)\widetilde{f}^{(a)}(z) \;,\\
	&\widetilde{d}^{(ba)}(z-w)\widetilde{e}^{(a)}(z)\widetilde{e}^{(b)}(w)\simeq(-1)^{|a||b|}\widetilde{d}^{(ba)}(z-w)\widetilde{e}^{(b)}(w)\widetilde{e}^{(a)}(z) \;,\\
	&\widetilde{d}^{(ab)}(z-w)\widetilde{f}^{(a)}(z)\widetilde{f}^{(b)}(w)\simeq(-1)^{|a||b|}\widetilde{d}^{(ab)}(z-w)\widetilde{f}^{(b)}(w)\widetilde{f}^{(a)}(z) \;,\\
	&\widetilde{\phi}^{a\Leftarrow b}(z-w-c)\widetilde{e}^{(a)}(z)\widetilde{f}^{(b)}(w)-(-1)^{|a||b|}\widetilde{f}^{(b)}(w)\widetilde{e}^{(a)}(z)\nonumber\\
	&\simeq \delta_{ab}\left(\delta(z-w-c)\widetilde{\psi}^{(a)}_+(w+c/2)-\delta(z-w+c)\widetilde{\psi}^{(a)}_-(z+c/2)-\delta(z-w)\widetilde{\omega}^{(a)}(z)\right)\;.
\end{align}
Here\footnote{This factor $\widetilde{d}^{(ab)}(z)$ is not redundant (which can be seen from the mode expansions) and is important when considering adding or removing atoms at non-generic positions in the crystal states. See Ref.~\refcite{Bao:2025hfu} for more details.},
\begin{equation}
	\widetilde{d}^{(ab)}(z):=\left(\prod_{I\in\{a\rightarrow b\}}(z+\epsilon_I)\right)\left(\prod_{I\in \{b\rightarrow a\}}(-z+\epsilon_I)\right)\;.
\end{equation}
Since this comes from the full 1-loop determinant (instead of ``half'' of it), it is in this sense called the double quiver Yangian/algebra.

For theories with 4 supercharges, the bond factors are always homogeneous. However, for theories with 2 supercharges, this is not necessarily the case. Therefore, to avoid the fractional powers in the current expansions, we may further rescale $\widetilde{\phi}^{(a)}(z)$ to $Z^{\frac{\mathfrak{t}}{2}(|\Lambda|-|\chi|)}\widetilde{\phi}^{a\Leftarrow b}(z)$. Moreover, we have chosen the $\mathbb{Z}_2$-grading such that $|a|\equiv|\Lambda|+1~(\text{mod}~2)$ for theories with 2 supercharges\footnote{This ensures that the Bose/Fermi statistics would not change under the dimensional reduction 4d $\mathcal{N}=1$ to 2d $\mathcal{N}=(0,2)$ for the corresponding currents although how the quiver algebras would be related is still not clear.}.

Although we have some algebra structures for both theories with 4 supercharges and those with 2 supercharges, we are not calling the double quiver algebras as the BPS algebras since many properties are still not clear. As mentioned in \S1, the quiver Yangians/BPS algebras have coproduct structures and have connections to vertex operators algebras and integrable systems. It is natural to expect these for the algebras for more general theories (although the stories of the vertex operator algebras and the Bethe/gauge correspondence themselves need to be developed as well). In the representations of the double quiver algebras, the extra information in the BPS counting is also encoded in the coefficients in the actions of the currents. It could be possible that there is a way to incorporate such information into the states of the representations. This might give rise to some new algebras that unveil the structures in the BPS counting and the connections to different areas.

\section*{Acknowledgments}
I am indebted to Rak-Kyeong Seong and Masahito Yamazaki for the collaboration on Ref.~\refcite{Bao:2024ygr} and to Masahito Yamazaki for the collaboration on Ref.~\refcite{Bao:2025hfu}. I would also like to thank the audience at the GTM seminar at Kavli IPMU for valuable questions and discussions. The research is supported by JSPS Grant-in-Aid for Scientific Research (Grant No.~23K25865).



\bibliographystyle{ws-ijmpa}
\bibliography{references}
\end{document}